\documentclass[10pt, journal]{IEEEtran}
\IEEEoverridecommandlockouts
\usepackage{cite}
\usepackage{amsfonts}
\usepackage{algorithmic}
\usepackage{float} 
\usepackage{url}
\usepackage{booktabs} 
\usepackage{siunitx} 
\usepackage{comment}
\usepackage{multirow} 
\usepackage{amsmath}
\usepackage{graphicx}
\usepackage{subcaption}
\usepackage{caption}       
\usepackage{hyperref}

\def\BibTeX{{\rm B\kern-.05em{\sc i\kern-.025em b}\kern-.08em
    T\kern-.1667em\lower.7ex\hbox{E}\kern-.125emX}}
    
\usepackage[acronym]{glossaries}

\usepackage{caption}

\usepackage{ifpdf}
\ifpdf
\else
\fi
\newacronym{psnr}{PSNR}{peak signal-to-noise ratio}
\newacronym{iot}{IoT}{internet of things}
\newacronym{awgn}{AWGN}{additive white Gaussian noise}
\newacronym{bpp}{BPP}{bits per pixel}
\newacronym{lic}{LIC}{learned image compression}
\newacronym{vqgan}{VQGAN}{vector quantized generative adversarial networks}
\newacronym{gan}{GAN}{generative adversarial network}
\newacronym{ssim}{SSIM}{structural similarity index measure}
\newacronym{hevc}{HEVC}{high efficiency video coding}
\newacronym{dct}{DCT}{discrete cosine transform}
\newacronym{vae}{VAE}{variational autoencoder}
\newacronym{amc}{AMC}{adaptive modulation and coding}
\newacronym{vqvae}{VQ-VAE}{vector quantized variational autoencoder}
\newacronym{snr}{SNR}{signal-to-noise ratio}
\newacronym{ber}{BER}{bit error rate}
\newacronym{svc}{SVC}{scalable video coding}
\newacronym{vq}{VQ}{vector quantization}
\newacronym{mse}{MSE}{mean squared error}

\makeatletter
\def\ps@IEEEtitlepagestyle{%
  \def\@oddhead{\parbox{\textwidth}{\centering \footnotesize ©2024 IEEE. This work has been submitted to the IEEE for possible publication. \\
  Copyright may be transferred without notice, after which this version may no longer be accessible.}\vspace{1em}}%
  \def\@oddfoot{}%
}
\makeatother

\begin{document}

\title{Deep Learning-Based Image Compression for Wireless Communications: Impacts on Reliability, Throughput, and Latency}

\author{
Mostafa~Naseri$^{\dagger}$ \quad  Pooya~Ashtari$^{\ddagger}$ \quad Mohamed~Seif$^{\mathsection}$ \quad Eli~De~Poorter$^{\dagger}$ \quad H.~Vincent~Poor$^{\mathsection}$ \quad Adnan~Shahid$^{\dagger}$ \\ [0.1in]
$^{\dagger}$ IDLab, Department of Information Technology at Ghent University - imec \\
$^{\ddagger}$ Department of Electrical Engineering (ESAT), STADIUS Center, KU Leuven \\
$^{\mathsection}$ Department of Electrical and Computer Engineering, Princeton University, Princeton
}

\maketitle

\begin{abstract} In wireless communications, efficient image transmission must balance reliability, throughput, and latency, especially under dynamic channel conditions. This paper presents an adaptive and progressive pipeline for learned image compression (LIC)-based architectures tailored to such environments. We investigate two state-of-the-art learning-based models: the hyperprior model and Vector Quantized Generative Adversarial Network (VQGAN). The hyperprior model achieves superior compression performance through lossless compression in the bottleneck but is susceptible to bit errors, necessitating the use of error correction or retransmission mechanisms. In contrast, the VQGAN decoder demonstrates robust image reconstruction capabilities even in the absence of channel coding, enhancing reliability in challenging transmission scenarios. We propose progressive versions of both models, enabling partial image transmission and decoding under imperfect channel conditions. This progressive approach not only maintains image integrity under poor channel conditions but also significantly reduces latency by allowing immediate partial image availability. We evaluate our pipeline using the Kodak high-resolution image dataset under a Rayleigh fading wireless channel model simulating dynamic conditions. The results indicate that the progressive transmission framework enhances reliability and latency while maintaining or improving throughput compared to non-progressive counterparts across various Signal-to-Noise Ratio (SNR) levels. Specifically, the progressive-hyperprior model consistently outperforms others in latency metrics, particularly in the 99.9th percentile waiting time—a measure indicating the maximum waiting time experienced by 99.9\% of transmission instances—across all SNRs, and achieves higher throughput in low SNR scenarios. where Adaptive WebP fails. Adaptive WebP, an extension of the WebP image format that dynamically adjusts compression parameters based on channel conditions, demonstrates superior image quality and throughput at higher SNRs when channel conditions are favorable. The proposed method proves particularly advantageous for Internet of Things (IoT) applications and task-oriented communications, where initial low-quality images suffice for immediate processing, and subsequent data refinement enhances performance as conditions permit.

\end{abstract}

\vspace{0.1in}
\begin{IEEEkeywords}
Image Compression, Wireless Communications, Learned Image Compression, Hyperprior Model, VQGAN, Progressive Transmission, Reliability, Throughput, Latency.
\end{IEEEkeywords}

\title{Learning-Based Image Compression for Wireless Communications: Impact on Reliability, Throughput, and Latency}
\section{Introduction}

\IEEEPARstart{E}{fficient} image transmission has become increasingly critical in modern wireless communication systems, driven by the rapid increase of applications that demand high-quality visual data even under challenging channel conditions. Applications such as remote sensing, autonomous driving, augmented reality, and \gls{iot} rely on the timely and reliable transmission of images to function effectively \cite{10614204, li2024exploiting}. These scenarios often operate in environments with limited bandwidth and high levels of interference, necessitating robust and efficient image compression techniques.

Traditional image compression methods, including JPEG2000 \cite{skodras2001jpeg2000}, WebP \cite{google_webp}, and BPG \cite{bellard_bpg}, are widely utilized due to their established performance and compatibility. However, these conventional techniques encounter significant limitations at extremely low bitrates. At such low bitrates, traditional compressors are compelled to employ large quantization steps, resulting in substantial degradation of image quality. Furthermore, these methods are highly sensitive to bit errors; even minor errors can render the entire image undecodable rather than merely degrading its quality. This sensitivity necessitates the use of additional error correction or error detection and retransmission mechanisms to ensure image integrity, thereby increasing latency and reducing overall system efficiency \cite{bourtsoulatze2019deep}.

Noisy wireless channels, particularly those affected by \gls{awgn}, intensify the challenges associated with image transmission. High bit error rates and limited bandwidth can severely degrade image quality, making it difficult to achieve acceptable performance at very low \gls{bpp}. While channel coding techniques can mitigate some errors, relying solely on them may not be sufficient or efficient for ultra-low bitrate image transmission. In such scenarios, achieving both high compression efficiency and robustness to noise becomes paramount, as traditional methods struggle to meet these dual requirements \cite{mao2024extreme}.

Recent advancements in learning-based compression and reconstruction methods offer promising solutions to these challenges. By leveraging deep learning architectures, these techniques can achieve significant compression while maintaining or even enhancing perceptual image quality \cite{cheng2020learned, balle2018variational, li2024exploiting}. \gls{lic} models, such as hyperprior models \cite{balle2018variational, liu2023learned} and \gls{vqgan} \cite{esser2021taming}, have demonstrated superior performance compared to traditional methods. While hyperprior models focus on quantization and lossless compression in the bottleneck of their autoencoders, \gls{vqgan} was originally developed as an image tokenizer for image generation transformers. Nonetheless, its utilization of vector quantization in the latent space facilitates robust image transmission even in the absence of channel coding, which is advantageous for real-time image transmission as it reduces latency and computational overhead associated with encoding and decoding processes. This characteristic makes \gls{vqgan} a valuable component in image transmission systems, especially under dynamic and poor channel conditions where traditional compression models may falter.

The integration of recent advancements in neural joint source-channel coding (JSCC) further underscores the progress and ongoing challenges in efficient image transmission over wireless channels. Notably, Yang \textit{et al.}~\cite{yang2024swinjscc} introduce a neural JSCC backbone based on the Swin Transformer architecture, which demonstrates remarkable adaptability to diverse channel conditions and transmission requirements. Their approach incorporates a code mask module that prioritizes channel importance, enabling adaptive transmission through a single, unified model. This is achieved by integrating the target rate into multiple layers of the image compressor. In contrast, our methodology leverages a hyperprior model that organizes feature maps in a sorted manner, facilitating a different form of adaptability. Additionally, while Yang \textit{et al.} focus primarily on the inference time of the encoder and decoder, our study focuses on the waiting time induced by dynamic channel conditions, providing a more comprehensive evaluation of system latency.

Furthermore, Wu \textit{et al.}~\cite{wu2024transformer} present JSCCformer-f, a wireless image transmission paradigm that capitalizes on feedback from the receiver to enhance transmission efficacy. The unified encoder in JSCCformer-f effectively utilizes semantic information from the source image, channel state information, and the decoder’s current belief about the source image derived from the feedback signal to generate coded symbols dynamically at each transmission block. However, similar to Yang \textit{et al.}, their work primarily addresses the inference time of the model without delving into the waiting time associated with image transmission. Moreover, JSCCformer-f necessitates continuous feedback from the decoder, which introduces additional communication overhead and complexity. Unlike their approach, our research focuses on progressive transmission and provides an in-depth analysis of transmission waiting time, an important factor in delay sensitive applications, eliminating the dependency on receiver feedback and thereby streamlining the transmission process.

These recent studies by Yang \textit{et al.}~\cite{yang2024swinjscc} and Wu \textit{et al.}~\cite{wu2024transformer} highlight significant strides in the development of flexible and efficient neural JSCC frameworks. However, they also reveal critical areas that remain unexplored, particularly concerning the comprehensive assessment of transmission latency and the optimization of progressive transmission strategies without relying on receiver feedback. Addressing these gaps, our work aims to enhance the robustness and efficiency of image transmission systems by focusing on both inference and transmission waiting times, thereby contributing to more reliable and low-latency wireless communication solutions.

Building upon these advancements, we propose a novel adaptive and progressive image transmission pipeline based on state-of-the-art \gls{lic} architectures. Specifically, we leverage the strengths of the hyperprior model and \gls{vqgan} to address the challenges of dynamic wireless channels by balancing reliability, throughput, and latency.

The hyperprior model is known for its exceptional compression performance due to effective quantization and lossless compression in the bottleneck of the autoencoder. However, it is highly sensitive to bit errors, which limits its application in noisy channels unless robust error correction or retransmission mechanisms are employed. Conversely, \gls{vqgan} utilizes vector quantization in the latent space, inherently providing robustness to bit errors and allowing the decoder to reconstruct images even without channel coding. This characteristic enhances reliability in noisy environments but may not achieve the same compression efficiency as the hyperprior model.

To harness the advantages of both models and overcome their individual limitations, we introduce progressive versions of these architectures. Our progressive transmission framework allows for partial image transmission and decoding, enabling immediate availability of coarse images under suboptimal channel conditions or limited throughput. As channel conditions improve or more bandwidth becomes available, additional data can be transmitted to progressively refine the image quality. This approach not only maintains image integrity under poor channel conditions but also significantly reduces latency by allowing immediate partial image availability.

We evaluate our proposed pipeline on the Kodak high-resolution image dataset, measuring performance in terms of \gls{psnr} and \gls{ssim}. Experimental results demonstrate that our progressive transmission framework significantly enhances reliability and reduces latency compared to non-progressive counterparts. Moreover, the progressive approach is particularly beneficial for \gls{iot} applications and task-oriented communications, where initial low-quality images are sufficient for immediate processing, and subsequent refinements improve performance as conditions permit.

By integrating adaptive and progressive \gls{lic} architectures, this work contributes to the advancement of intelligent communications by providing a robust and efficient solution for real-time image transmission in wireless systems. Our approach addresses the critical need for reliable, low-latency image delivery in environments with dynamic and challenging channel conditions, thereby supporting the demands of real-time computer vision tasks and other emerging applications.

In this work, we present the following key contributions to the field of efficient and robust image transmission over wireless channels:

\begin{enumerate}
    \item \textbf{First Progressive Transmission Framework Based on \gls{lic}:} We introduce a novel progressive transmission pipeline tailored for hyperprior-based \gls{lic} architectures. To the best of our knowledge, this is the first approach that leverages \gls{lic} for progressive image transmission, enabling efficient and adaptable transmission under varying channel conditions.

    \item \textbf{Use of Residual Vector Quantization with \gls{vqgan}:} This work is the first to employ residual vector quantization within a \gls{vqgan}-based framework for progressive image transmission. Our method ensures reliable image reconstruction in noisy channels without relying on traditional channel coding. Additionally, the progressive strategy facilitates immediate partial image availability, significantly reducing latency for real-time applications.

    \item \textbf{Analysis of Waiting Time and Practical Implementation:} We conduct an analysis of waiting time in progressive image transmission, addressing an aspect that previous works have overlooked by focusing primarily on inference time. This analysis provides deeper insights into the real-time performance and responsiveness of our transmission framework. Furthermore, by utilizing readily available image encoders, we enhance the reproducibility and ease of further development of our work, making it more accessible for future research and practical applications.

\end{enumerate}

These contributions collectively advance the state-of-the-art in intelligent image transmission over wireless channels, providing robust, low-latency, and efficient solutions tailored to the demands of modern and emerging applications\url{}\footnote{Project code: \url{https://github.com/M0574F4/LIC_TX}}.

\section{Related Work}

\subsection{Traditional Image Compression Methods}
Traditional image compression techniques have been the cornerstone of digital image transmission for decades. Standards such as JPEG \cite{125072}, WebP \cite{google_webp}, and \gls{hevc} \cite{sullivan2012overview} are widely adopted due to their balance between compression efficiency and computational complexity. JPEG, one of the earliest compression standards, employs the \gls{dct} to reduce spatial redundancy, followed by quantization and entropy coding \cite{125072}. Despite its widespread use, JPEG suffers significant artifacts and quality degradation at extremely low bitrates due to large quantization steps, which discard essential image details \cite{jiang2021towards}.

WebP, developed by Google, extends the capabilities of JPEG by incorporating both lossy and lossless compression techniques, leveraging predictive coding and entropy encoding to achieve better compression ratios \cite{google_webp}. \gls{hevc} further improves compression efficiency by using advanced techniques such as block partitioning, motion compensation, and in-loop filtering, making it suitable for high-resolution and high-dynamic-range images \cite{sullivan2012overview}. However, these traditional methods are highly sensitive to bit errors, especially in noisy wireless channels. Minor bit errors can lead to significant artifacts or render the entire image undecodable, necessitating the use of robust error correction or retransmission strategies to maintain image integrity \cite{bourtsoulatze2019deep}.

\subsection{Learned Image Compression Methods}
Recent advancements in deep learning have encouraged the development of \gls{lic} models, which leverage neural networks to outperform traditional compression standards in both compression efficiency and adaptability. hyperprior models, introduced by Ballé et al. \cite{balle2018variational}, utilize a hypernetwork to model the distribution of latent representations, enabling more efficient entropy coding and improved compression performance. These models achieve state-of-the-art results by optimizing rate-distortion trade-offs through end-to-end training \cite{liu2023learned}.

\gls{vqgan}, originally designed as image tokenizers for generative transformers \cite{esser2021taming}, has been repurposed for image compression by leveraging vector quantization in the latent space of autoencoder architectures \cite{mao2024extreme}. Unlike traditional \gls{lic} models, \gls{vqgan} is not inherently a compression model but offers robust image reconstruction capabilities where errors in the tokenized latent representation result in only localized quality degradation rather than making the entire image undecodable. However, \gls{vqgan}'s primary focus on image generation rather than compression introduces challenges in achieving optimal compression ratios without compromising image quality.

Other notable \gls{lic} approaches include \glspl{vae} \cite{zhou2018variational} and \glspl{gan} \cite{agustsson2019generative}, which have been tailored for compression by incorporating probabilistic models and adversarial training to enhance perceptual quality. These models demonstrate significant improvements over traditional methods, particularly in maintaining image fidelity at lower bitrates.

\subsection{Progressive Transmission in Wireless Communications}
Progressive transmission techniques have gained traction as a means to enhance image transmission reliability and reduce latency in wireless communications. Progressive compression schemes, such as \gls{svc} \cite{911159} and layered image compression \cite{jia2019layered}, enable the transmission of image data in multiple layers, allowing receivers to reconstruct low-quality images quickly and progressively enhance them as more data becomes available. This approach is particularly beneficial in scenarios with fluctuating channel conditions, as it ensures that partial data can be utilized effectively without waiting for complete transmission \cite{lan2022progressive}.

In the context of wireless communications, progressive transmission aligns well with the need for balancing reliability, throughput, and latency. Techniques such as unequal error protection \cite{borade2009unequal} and \gls{amc} \cite{Goldsmith_2005} have been employed to dynamically adjust transmission parameters based on channel quality. However, integrating progressive transmission with \gls{lic} models remains a relatively unexplored area, presenting opportunities for enhancing image transmission robustness and efficiency.

To the best of our knowledge, progressive image transmission tailored specifically for learned image compression has not been extensively investigated. However, researchers outside the communications domain have explored progressive decoding strategies. For instance, \cite{hojjat2023progdtd} developed a progressive decoder, demonstrating the feasibility of controllable rates in an autoencoder with a hyperprior branch. Additionally, Flowers et al. \cite{flowers2024bric} employed a \gls{vqvae} model and proposed a hierarchical architecture that performs residual vector quantization in the bottleneck of the image autoencoder. These studies highlight the potential of progressive decoding and hierarchical quantization approaches, inspiring our proposed adaptive and progressive image transmission pipeline integrated with \gls{lic} models for wireless environments.

\section{Background and Preliminaries}

Efficient image transmission in wireless communications necessitates a comprehensive understanding of both the communication channel characteristics and the fundamentals of image compression techniques. This section provides the essential background required to appreciate the challenges and innovations presented in this study.

\subsection{System Model}

The performance of image transmission systems in wireless environments is intrinsically linked to the properties of the communication channels. To design robust image compression and transmission schemes, it is crucial to model these channels accurately while maintaining a balance between complexity and practicality.

\subsubsection{Wireless Channel Model}
\label{sec:channel}
A wireless communication channel can be mathematically modeled as:
\begin{equation}
y = hx + n,
\end{equation}
where:
\begin{itemize}
    \item $y \in \mathbb{C}^L$ is the received signal vector,
    \item $h \in \mathbb{C}^L$ represents the channel coefficients,
    \item $x \in \mathbb{C}^L$ is the transmitted signal vector,
    \item $n \sim \mathcal{CN}(0, \sigma^2I)$ is the \gls{awgn} vector with zero mean and covariance matrix $\sigma^2I$.
\end{itemize}
and $L$ is dimensionality of the signal vectors.
In this model, $h$ encompasses both large-scale fading (path loss, shadowing) and small-scale fading (multipath effects). The \gls{awgn} component $n$ captures the thermal noise inherent in the communication system.

This formulation allows us to abstract the complex channel characteristics into an effective transmission rate $R$, which is crucial for designing adaptive and progressive transmission schemes.

\subsubsection{Capacity Analysis under Finite Block Length Constraints}

Shannon's capacity theorem provides a fundamental limit on the maximum achievable transmission rate $C$ for a given channel, ensuring reliable communication. However, Shannon's theorem assumes infinitely long codewords, which is impractical for real-world applications where finite block lengths $n$ are used. The finite block length capacity introduces a trade-off between throughput and delay:
\begin{equation}
R = C - \sqrt{\frac{V}{n}}Q^{-1}(\epsilon),
\end{equation}
where:
\begin{itemize}
    \item $R$ is the achievable rate,
    \item $V$ is the channel dispersion,
    \item $n$ is the block length,
    \item $Q^{-1}(\epsilon)$ is the inverse of the Q-function evaluated at the error probability $\epsilon$.
\end{itemize}

This relationship highlights that shorter block lengths, which are desirable for low latency, result in rates $R$ that are below the Shannon capacity $C$. Consequently, there is a trade-off between maximizing throughput and minimizing delay, which is particularly relevant for designing progressive and adaptive transmission schemes where latency is a critical factor \cite{9635675}.

For the purposes of this study, we use achievable transmission rate $R$, which is a function of the available bandwidth $B$ and the \gls{snr}. According to Shannon's capacity theorem \cite{shannon1948mathematical}, the channel capacity $C$ in bits per second (bps) is given by:
\begin{equation}
C = B \log_2\left(1 + \frac{P}{N_0 B}\right),
\end{equation}
where:
\begin{itemize}
    \item $P$ is the transmit power,
    \item $N_0$ is the noise power spectral density.
\end{itemize}
\subsubsection{Performance Criteria} We consider three criteria based on system parameters and channel conditions: throughput, reliability, and latency, which we will describe in full detail next.


\paragraph{Throughput}

Throughput refers to the effective data transmission rate and is constrained by the channel capacity $C$. Typically, in communication systems, throughput is measured in bits per second (bps) to indicate how much data can be transmitted within a given time frame. However, in this work, which focuses on computer vision tasks, we define throughput as pixels per second, as this metric more directly relates to the amount of visual data processed over time. Higher throughput, measured in pixels per second, enables faster image processing and transmission, which is essential for applications requiring real-time or near-real-time data delivery.

\paragraph{Reliability}
Reliability pertains to the accuracy and integrity of the transmitted image data. It is influenced by factors such as the \gls{ber}. In task-oriented communications, reliability is defined not only by the \gls{ber} but also by the successful completion of specific tasks, such as object detection or recognition, based on the received images. However, in this work, where specific tasks are not the primary focus, reliability is primarily assessed using metrics like \gls{psnr} and \gls{ssim}. These metrics provide a quantifiable measure of image fidelity and integrity, effectively capturing reliability in the absence of task-specific evaluations.

\paragraph{Latency}

Latency $L$ is the time delay between the initiation of image transmission and its successful reception and reconstruction at the receiver. End-to-end latency can be decomposed into several components:
\begin{equation}
L = L_c + L_t + L_d,
\end{equation}
where:
\begin{itemize}
    \item $L_c$ is the compression latency,
    \item $L_t$ is the transmission latency,
    \item $L_d$ is the decompression latency.
\end{itemize}
Compression latency ($L_c$) is the time taken to compress the image before transmission. Transmission latency ($L_t$) is the time required to transmit the compressed data over the wireless channel, which is influenced by the available transmission rate and the size of the data. Decompression latency ($L_d$) is the time taken to decompress and reconstruct the image at the receiver.

Progressive transmission primarily aims to reduce the transmission latency ($L_t$) by enabling the receiver to reconstruct a low-quality version of the image quickly, followed by incremental enhancements as more data becomes available. This approach minimizes the perceived latency, allowing for immediate partial image availability at the cost of initially lower image quality.

\subsection{Learned Image Compression Fundamentals}

\gls{lic} leverages deep learning techniques to surpass traditional compression methods in terms of compression efficiency and adaptability to varying channel conditions. \gls{lic} models typically employ neural network architectures that are trained end-to-end to optimize the trade-off between compression rate and image quality.

\subsubsection{Overview of Autoencoder Architectures}

At the heart of most \gls{lic} models lies the autoencoder architecture, which comprises an encoder, a bottleneck (latent space), and a decoder \cite{balle2018variational}. The encoder transforms the input image $I$ into a compact latent representation $z$, which is then quantized and compressed for transmission. The decoder reconstructs the image $\hat{I}$ from the compressed latent code:
\begin{equation}
z = \text{Quantize}(\text{Encoder}(I)),
\end{equation}
\begin{equation}
\hat{I} = \text{Decoder}(z).
\end{equation}

The autoencoder framework is highly flexible and can be enhanced with various mechanisms to improve compression efficiency and image quality. For instance, incorporating attention mechanisms or residual connections can enhance the model's ability to capture intricate image details \cite{liu2023learned}. Additionally, probabilistic models within the autoencoder allow for better entropy modeling, leading to more efficient compression \cite{mentzer2018conditional}.

\subsubsection{Vector Quantization in Latent Spaces}

\gls{vq} is a technique used to discretize the continuous latent representations produced by the encoder. In the context of \gls{lic}, \gls{vq} reduces the dimensionality of the latent space, facilitating more efficient compression \cite{8798635}. For example, \gls{vqgan} utilize \gls{vq} in their autoencoder architectures to create discrete latent codes that are easier to compress and transmit \cite{esser2021taming}.

The integration of \gls{vq} in \gls{lic} models without lossless compression offers several advantages. Discrete latent spaces are inherently more resilient to transmission noise, enhancing the reliability of image reconstruction \cite{santhirasekaram2022vector}. Additionally, hierarchical \gls{vq} approaches, such as residual vector quantization, enable the progressive refinement of image details by encoding residual information at multiple levels \cite{flowers2024bric}. This hierarchical structure not only improves compression efficiency but also supports progressive transmission by allowing partial data to enhance image quality incrementally. Furthermore, recent advancements have explored the use of transformer architectures in conjunction with \gls{vq}-based models for generating high-fidelity images from discrete latent codes, bridging the gap between compression and image generation \cite{esser2021taming}.

These features make \gls{vq}-based \gls{lic} models particularly well-suited for adaptive and progressive image transmission in dynamic wireless environments.

\section{Proposed Methodology}

This section presents our novel adaptive and progressive image transmission pipeline tailored for learned image compression (LIC)-based architectures. We explore two state-of-the-art LIC models: a hyperprior-based model and a \gls{vqgan}. Our proposed pipeline is designed to operate efficiently under dynamic channel conditions, balancing reliability, throughput, and latency.

\subsection{System Architecture Overview}

\begin{figure*}[t]
    \centering
    \includegraphics[width=0.9\textwidth]{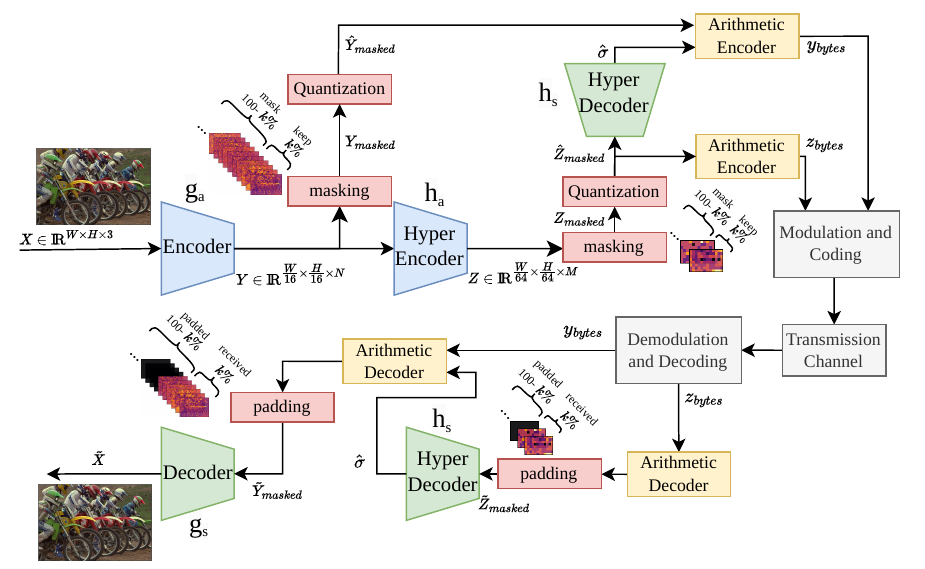}
    \caption{\small{System model for transmission based on the hyperprior model, illustrating the retention of $k\%$ of feature maps while masking the remainder in $Y$ and $Z$ at the transmitter, and the reconstruction by padding zeroes in the masked $100 - k\%$ feature maps at the receiver. The selection of $k$ is based on channel conditions and service requirements.
}}
    \label{fig:sys_model_hyperprior}
\end{figure*}


Our transmission pipeline is structured around two primary LIC models: the hyperprior-based model and the \gls{vqgan}-based model. Each model's architecture is illustrated in Figs. \ref{fig:sys_model_hyperprior} and \ref{fig:vq_sys_model}.

\subsubsection{Hyperprior-Based Model Integration}

The hyperprior-based model utilizes a variational autoencoder structure that enhances compression efficiency through the use of a hyperprior. The architecture of this model, depicted in Fig. \ref{fig:sys_model_hyperprior}, employs the following mathematical formulation \cite{balle2018variational}:

\begin{align}
    Y &= g_a(X), \\
    Y_{\text{masked}} &= \text{Mask}(Y, k\%), \\
    \hat{Y}_{\text{masked}} &= \text{Quantize}(Y_{\text{masked}}), \\
    Z &= h_a(Y), \\
    Z_{\text{masked}} &= \text{Mask}(Z, k\%), \\
    \hat{Z}_{\text{masked}} &= \text{Quantize}(Z_{\text{masked}}), \\
    \hat{\sigma} &= h_s(\hat{Z}_{\text{masked}}), \\
    y_{\text{bytes}} &= \text{ArithmeticEncode}(\hat{Y}_{\text{masked}}, \hat{\sigma}), \\
    z_{\text{bytes}} &= \text{ArithmeticEncode}(\hat{Z}_{\text{masked}}),
\end{align}
where,
\begin{itemize}
    \item $g_a$ is the encoder that transforms the input image $X \in \mathbb{R}^{W \times H \times 3}$ into a latent representation $Y \in \mathbb{R}^{\frac{W}{16} \times \frac{H}{16} \times N}$.
    \item $\text{Mask}(Y, k\%)$ masks $k\%$ of the channels in $Y$, preserving the remaining channels.
    \item $\text{Quantize}(Y_{\text{masked}})$ quantizes the masked latent representation to produce $\hat{Y}_{\text{masked}}$.
    \item $h_a$ is the hyperprior encoder that processes $Y$ to generate hyper latent $Z \in \mathbb{R}^{\frac{W}{64} \times \frac{H}{64} \times M}$.
    \item $\text{Mask}(Z, k\%)$ and $\text{Quantize}(Z_{\text{masked}})$ follow similar processing steps for the hyper latent.
    \item $h_s$ decodes $\hat{Z}_{\text{masked}}$ to estimate the scale parameter $\hat{\sigma}$, crucial for arithmetic encoding.
    \item $\text{ArithmeticEncode}$ generates byte streams $y_{\text{bytes}}$ and $z_{\text{bytes}}$ using the quantized and masked representations.
\end{itemize}

At the receiver end, depicted in Fig. \ref{fig:sys_model_hyperprior}, the decoding process unfolds as follows:

\begin{align}
    \tilde{Z}_{\text{masked}} &= \text{ArithmeticDecode}(z_{\text{bytes}}), \\
    \hat{\sigma} &= h_s(\tilde{Z}_{\text{masked}}), \\
    \tilde{Y}_{\text{masked}} &= \text{ArithmeticDecode}(y_{\text{bytes}}, \hat{\sigma}), \\
    \tilde{X} &= g_s(\tilde{Y}_{\text{masked}}),
\end{align}
where:
\begin{itemize}
    \item $g_s$ is the decoder that reconstructs the image $\tilde{X}$ from the processed $\tilde{Y}_{\text{masked}}$.
\end{itemize}

During training, a random selection of $k\%$ of the feature maps is masked to train the network to focus on significant features where $k = 100 \times u$ and $u \sim \text{Uniform}(0,1)$. During inference, the selection of $k\%$ is dynamically adjusted based on real-time channel conditions and application requirements, enabling adaptive transmission efficiency. This adaptive selection allows the system to decode the image from a minimal number of feature maps initially, with subsequent transmissions providing additional feature maps to enhance image quality progressively.

\subsubsection{\gls{vqgan}-Based Model Integration}

The \gls{vqgan}-based model employs vector quantization within its generative adversarial network architecture. This model's system architecture is illustrated in Fig. \ref{fig:vq_sys_model}, with the following process flow:
\begin{figure*}[t]
    \centering
    \includegraphics[width=0.9\textwidth]{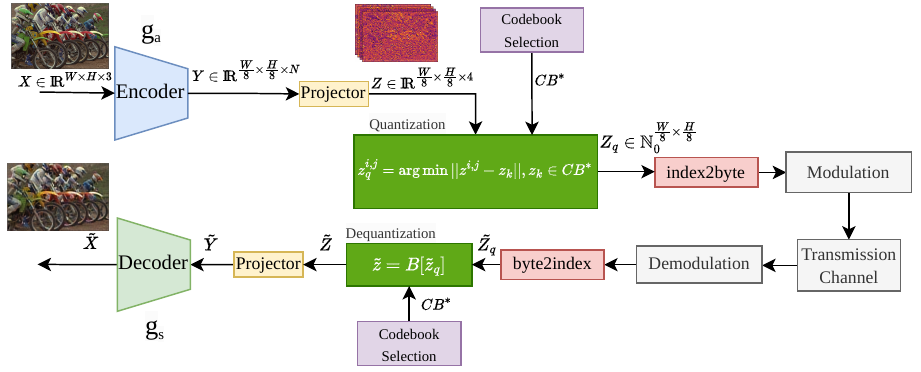}
    \caption{\small{System model for \gls{vqgan}-based image transmission, illustrating the generation of feature maps ($Y$) and token maps ($Z$) during the encoding and decoding process.
}}
    \label{fig:vq_sys_model}
\end{figure*}
\begin{align}
    Y &= g_a(X), \\
    Z &= P_E(Y), \\
    Z_q &= \text{VectorQuantize}(Z, CB), \\
    \tilde{Z} &= \text{Dequantize}(\tilde{Z}_q, CB), \\
    \tilde{Y} &= P_D(\tilde{Z}), \\
    \tilde{X} &= g_s(\tilde{Y}),
\end{align}
where,
\begin{itemize}
    \item $g_a$ is the encoder transforming the input image $X$ into the latent representation $Y \in \mathbb{R}^{\frac{W}{8} \times \frac{H}{8} \times N}$.
    \item $P_E$ projects $Y$ into a lower-dimensional space $Z \in \mathbb{R}^{\frac{W}{8} \times \frac{H}{8} \times 4}$.
    \item $\text{VectorQuantize}(Z, CB)$ quantizes $Z$ into the token map $Z_q$, using the codebook $CB$.
\end{itemize}

Additionally, the \gls{vqgan}-based model incorporates residual codebook clustering, enabling a hierarchical architecture that performs residual vector quantization in the bottleneck of the image autoencoder. This hierarchical structure allows for progressive decoding by transmitting indices corresponding to coarse and fine codebooks in successive transmissions, thereby refining the image quality incrementally without relying on lossless compression.

\subsection{Adaptive and Progressive Transmission Mechanism}

Our progressive transmission mechanism is designed to adaptively adjust the amount of data transmitted based on real-time channel conditions and application-specific requirements. This mechanism is integral to both the hyperprior-based and \gls{vqgan}-based models, enabling them to balance reliability, throughput, and latency effectively.

\subsubsection{Adaptive Pipeline Design}

Rather than redesigning the communication system \cite{bourtsoulatze2019deep}, our approach leverages existing systems where modulation and coding schemes are dynamically selected based on channel condition. This adaptability allows our pipeline to be seamlessly integrated into various communication infrastructures.

In each transmission slot, a permissible bit budget $N_{\text{bits}}$, determined by the current channel rate—which is influenced by factors such as bandwidth and noise level—is allocated. Based on this $N_{\text{bits}}$, the system performs the following selections:

Hyperprior-Based Model: Selects the maximum number of feature maps such that the top $k\%$ of channels fit within the allocated bit budget.

\gls{vqgan}-Based Model: Chooses the number of relevant codebooks for vector quantization based on the available channel rate.

\subsubsection{Progressive Image Decoding}

\paragraph{Hyperprior-Based Model}

In this experiment, we analyzed the sensitivity of the Learned Image Compression-TCM (LIC-TCM) model from Liu et al. \cite{liu2023learned} to channel masking. Specifically, we initiated the masking of a percentage of channels in the bottleneck and observed non-uniform sensitivity across different channels. Masking certain channels resulted in significant PSNR degradation, while others had minimal impact.

To further investigate, we treated the problem as a feature pruning task. We performed inference on 100 randomly selected batches of images (batch size = 8) from the ImageNet dataset, systematically masking each feature map and recording the Mean Squared Error (MSE) degradation in image reconstruction. By averaging the MSE degradation across all tested images, we derived an importance metric for each feature map.

\begin{figure}[t]
    \centering
    \includegraphics[width=0.85\linewidth]{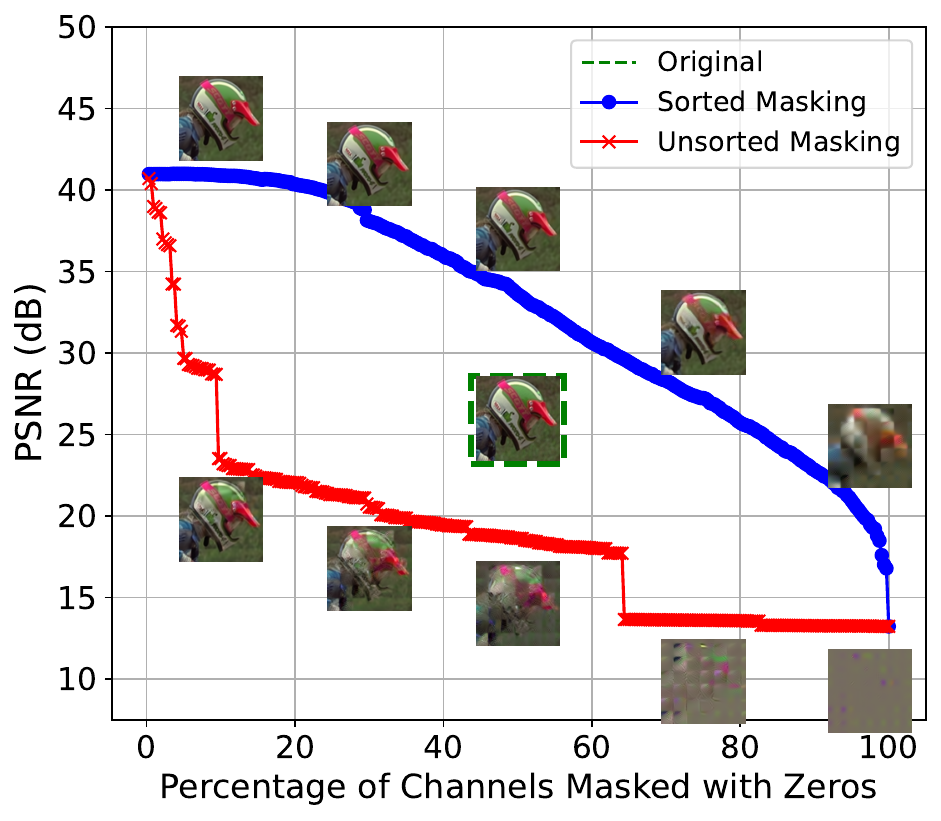}
    \caption{\small{PSNR degradation versus the percentage of masked bottleneck channels for the image \textit{kodim05}. The solid line represents unsorted masking, while the dashed line corresponds to observer-based feature masking.}}
    \label{fig:psnr_drop}
\end{figure}

Our results demonstrated that masking channels in order of least to most importance based on the averaged MSE degradation achieved a smooth PSNR drop and maintained better image reconstruction quality across varying mask percentages, as illustrated in Figure \ref{fig:psnr_drop}. This finding motivates our progressive image transmission approach for hyperprior-based models. We utilize the sorting mechanism from Hojjat et al. \cite{hojjat2023progdtd}, which ranks feature maps according to their importance metrics, allowing for an adaptive and progressive transmission strategy that prioritizes critical feature maps under varying channel conditions.

At the receiver, the progressive transmission is handled by maintaining the previously received $\tilde{Y}_{\text{masked}}$ feature maps. With each new transmission, additional feature maps are received and concatenated with the existing ones, allowing the decoder to incrementally refine the reconstructed image $\tilde{X}$. This approach ensures that a low-quality version of the image is available quickly, with subsequent transmissions enhancing the image quality as more data becomes available.

\paragraph{\gls{vqgan}-Based Model}

For the \gls{vqgan}-based model, we adopt an advanced hierarchical vector quantization approach to facilitate progressive image transmission. Building upon the methodology proposed by Zhu et al. \cite{zhu2024scaling}, we use their projector to compress the embedding dimension to 4, thereby reducing the latent space complexity. Subsequently, it expands the codebook size to 100K, establishing a large codebook that serves as the foundation for our residual vector quantization-based progressive transmission.

To manage this extensive codebook efficiently, we perform two types of clustering:

K-Means-Based Clustering: We utilize Facebook AI Similarity Search (Faiss) \cite{douze2024faiss} to perform k-means clustering, partitioning the large codebook into smaller, manageable subsets corresponding to different bits per index ($bpi$). The clustering is configured with a variable number of clusters, specifically $2^{bpi}$, to represent different reconstruction qualities. For each desired $bpi$, the algorithm clusters the embeddings into $2^{bpi}$ clusters. These clustered codebooks are then extracted and stored, facilitating efficient codebook selection based on the available bit budget $N_{\text{bits}}$ during transmission.

The codebook selection process is formalized as an optimization problem:
\begin{equation}
    CB^* = \arg\max_{B} \quad bpi \quad \text{s.t.} \quad n_{\text{bits}} \leq N_{\text{bits}},
\end{equation}
where the number of bits $n_{\text{bits}}$ is defined as:
\begin{equation}
    n_{\text{bits}} = \left(\frac{W}{8}\right) \left(\frac{H}{8}\right) bpi,
\end{equation}
with $B^*$ being the selected codebook among all available codebooks, $CB$, $W \times H$ denoting the dimensions of the image, and 8 being the down sampling factor across each spatial dimension of the encoder. Here, $bpi$ represents the number of bits per index corresponding to the selected codebook. This formulation ensures that the selected codebook maximizes the bits per index without exceeding the transmission bit budget.

Residual Quantizer (RQ): For progressive transmission, we implement a residual quantizer (RQ) that quantizes the residuals of the input vectors sequentially. At each encoding stage $m$, the RQ selects the codeword $c_m$ that best approximates the residual of the input vector $x$ relative to the previously encoded steps:
\begin{equation}
    c_m = \underset{j=1,\dots,K}{\text{argmin}} \left\| \sum_{i=1}^{m-1} T_i[c_i] + T_m[j] - x \right\|^2,
\end{equation}
where $T_i[c_i]$ represents the codeword from the $i^{\text{th}}$ transmission stage indexed by $c_i$, and $K$ is the number of codewords in the codebook. This sequential encoding allows the receiver to reconstruct the image progressively by first decoding coarse details and then refining them with finer residuals as more indices are received.

\subsubsection{Progressive Transmission Strategy}

In the \gls{vqgan}-based model, the receiver maintains a list of previously decoded indices corresponding to coarse codebooks. As new indices referencing finer codebooks are received, they are integrated with the existing indices to progressively decode and enhance the image. This hierarchical decoding process enables the immediate reconstruction of a basic image from the coarse indices, followed by incremental improvements as additional fine-grained indices are received.

To optimize progressive transmission, we dynamically determine the number of encoding stages that can be transmitted within the available bit budget $N_{\text{bits}}$. Given that each stage of residual quantization requires a fixed number of bits $n_{\text{bits}}$, the maximum number of stages $M_{\text{stages}}$ that can be accommodated is calculated as:
\begin{equation}
    M_{\text{stages}} = \min\left(\left\lfloor \frac{N_{\text{bits}}}{n_{\text{bits}}} \right\rfloor, M_{\text{stages}}^{\text{max}}\right),
\end{equation}
where $\left\lfloor \cdot \right\rfloor$ denotes the floor operation and $M_{\text{stages}}^{\text{max}}$ is the maximum number of codebooks learned for the residual vector quantization. This ensures that the total number of bits used does not exceed the budget $N_{\text{bits}}$.

Each encoding stage corresponds to a residual codebook that quantizes the residuals of the input vector relative to the previous stages. All residual codebooks are of the same size, and each stage incrementally refines the image quality by addressing the residual errors from earlier stages. By determining $M_{\text{stages}}$, we can transmit up to $M_{\text{stages}}$ stages of residual quantization within the given bit budget, allowing the receiver to progressively enhance the image quality as more data becomes available.

This adaptive selection of encoding stages based on the bit budget ensures that the most critical residuals are transmitted first, facilitating a balanced trade-off between image quality and transmission efficiency. Consequently, our approach leverages residual vector quantization to enhance the fidelity of the reconstructed image progressively, enabling robust and scalable image transmission under varying channel conditions without relying on lossless compression.

Using adaptive and progressive image transmission pipelines that integrate hyperprior-based and \gls{vqgan}-based LIC models to address the challenges of dynamic wireless channels by dynamically adjusting the transmitted data based on real-time channel conditions and employing progressive decoding strategies, we effectively balance reliability, throughput, and latency.

\begin{table}[h]
    \centering
    \caption{Notations}
    \label{tab:notations}
    \begin{tabular}{ll}
        \toprule
        \textbf{Symbol} & \textbf{Description} \\
        \midrule
        $R$ & Achievable transmission rate (bps) \\
        $B$ & Available bandwidth (Hz) \\
        $\text{SNR}$ & Signal-to-noise ratio \\
        $C$ & Channel capacity (bps) \\
        $P$ & Transmit power (W) \\
        $N_0$ & Noise power spectral density (W/Hz) \\
        $L$ & Latency (seconds) \\
        $L_c$ & Compression latency (seconds) \\
        $L_t$ & Transmission latency (seconds) \\
        $L_d$ & Decompression latency (seconds) \\
        $\text{BER}$ & Bit error rate \\
        $\text{PSNR}$ & Peak signal-to-noise ratio \\
        $\text{SSIM}$ & Structural similarity index measure \\
        $h \in \mathbb{C}^L$ & Channel coefficients \\
        $X$ & input image\\
        $Y$ & Latent representation of $X$\\
        $Y_{\text{masked}}$ & Masked latent representation $Y$ \\
        $\hat{Y}_{\text{masked}}$ & Quantized masked latent representation \\
        $Z$ & Hyper latent \\
        $Z_{\text{masked}}$ & Masked hyper latent $Z$ \\
        $\hat{Z}_{\text{masked}}$ & Quantized masked hyper latent $Z$ \\
        $\hat{\sigma}$ & Estimated scale parameter\\
        $\tilde{Z}_{\text{masked}}$ & Decoded masked hyper latent $Z$ \\
        $\tilde{Y}_{\text{masked}}$ & Decoded masked latent representation $Y$ \\
        $\tilde{X}$ & Reconstructed image\\
        $CB^*$ & Selected codebook \\
        $M_{\text{stages}}^{\text{max}}$ & Maximum number of codebooks in RQ \\

        \bottomrule
    \end{tabular}
\end{table}

\section{Experimental Setup}

To evaluate the performance of our proposed adaptive and progressive image transmission pipeline, we conducted comprehensive experiments designed to assess the efficacy of our models under realistic wireless communication conditions. This section details the experimental setup, including the datasets used, model configurations, simulation parameters, implementation environment, and evaluation metrics.
\subsection{Dataset and Models}

\subsubsection{Dataset}
We utilized the widely recognized Kodak dataset \cite{kodak_dataset} for evaluating our models.
Kodak Dataset Consists of 24 high-quality, uncompressed color images frequently used as a benchmark for image compression and transmission evaluations. The images cover a variety of scenes and are of size $768 \times 512$ or $512 \times 768$ pixels.

All images were preprocessed to match the input requirements of the pretrained models, normalization. No data augmentation techniques were applied during testing to maintain consistency.

\subsubsection{Models}

Our study investigates two primary \gls{lic} models: a hyperprior-based model and a \gls{vqgan}-based model.

\paragraph{Hyperprior-Based Model}

We employed the architecture described in \cite{hojjat2023progdtd}, which is based on a hyperprior variational autoencoder for image compression. The model configurations are as follows:

\begin{itemize}
    \item Latent Channels ($N_Y$): 192 channels
    \item Hyper Latent Channels ($N_Z$): 128 channels
\end{itemize}
and is trained using rate-distortion parameter $\lambda = 0.1$, and for using 0-100$\%$ of the channels.
This particular model is 
\paragraph{VQGAN-Based Model}

For the \gls{vqgan}-based model, we utilized the LIC architecture proposed in \cite{zhu2024scaling}, which integrates vector quantization with GANs for image compression. The model specifications include:

\begin{itemize}
    \item Codebook: A large initial codebook of 100,000 entries.
    \item Projection: A projector that compresses the embedding dimension to 4, facilitating efficient vector quantization.
\end{itemize}

We use  pretrained models on the ImageNet dataset \cite{deng2009imagenet}, and no additional training was conducted as part of this study.
\paragraph{Adaptive WebP Model}

In addition to the learned image compression models, we utilize an adaptive WebP \cite{google_webp}, which dynamically delivers the highest quality image within a specified range of quality factors based on channel conditions. The adaptive WebP approach adjusts compression parameters in real-time to optimize the balance between image quality and bitrate.

\subsection{Implementation Environment}

\subsubsection{Hardware}

All experiments were conducted on a high-performance computing setup equipped with:

\begin{itemize}
    \item Processor: Intel(R) Xeon(R) Silver 4310 CPU @ 2.10GHz with 48 cores.
    \item Graphics Processing Unit (GPU): NVIDIA A40 with 46,068 MiB memory, supporting accelerated computations for deep learning models.
\end{itemize}

\subsubsection{Software}

The software environment was configured as follows:

\begin{itemize}
    \item Operating System: Ubuntu 20.04 LTS.
    \item Deep Learning Framework: PyTorch version 2.2.1.
    \item CUDA: Version 12.2, enabling GPU acceleration.
    \item FAISS \cite{douze2024faiss}: Utilized for efficient k-means clustering during codebook generation.
    \item Additional Libraries: NumPy, SciPy, and other standard scientific computing libraries.
\end{itemize}

\subsection{Channel Simulation}

To model realistic wireless communication environments characterized by multipath propagation and rapid signal attenuation, we employed a Rayleigh fading channel model. This model is widely used to represent multipath fading in wireless communications. The channel coefficients were generated using the sum-of-sinusoids method, approximating Clarke's model for flat fading channels \cite{xiao2006novel}. The key parameters of the channel simulation are as follows:

\begin{itemize}
    \item Maximum Doppler Frequency ($f_d$): 10 Hz, representing a moderate level of mobility.
    \item Symbol Duration ($T_s$): 1 ms.
    \item Bandwidth ($B$): 100 kHz.
    \item \gls{snr}: Varied over \{-10, -5, 0, 5\} dB to simulate different channel conditions.
\end{itemize}

We are mainly interested in image transmission in challenging channel conditions hence low \gls{snr}s and limited channel bandwidth are assumed. To ensure statistical reliability and average out the randomness inherent in Rayleigh fading channels, we simulated 1,000 independent channel realizations.

\subsection{Performance Metrics}

Evaluating the performance of image compression and transmission systems involves multiple metrics that assess various aspects of system efficiency and image quality. The key metrics considered in this study are \gls{psnr}, \gls{ssim}, throughput, and latency.

\subsubsection{Image Quality Metrics}

\gls{psnr} is a widely used objective metric for measuring the quality of reconstructed images compared to the original images. It is defined as:
\begin{equation}
\text{PSNR} = 10 \cdot \log_{10} \left( \frac{\text{MAX}^2}{\text{MSE}} \right),
\end{equation}
where $\text{MAX}$ represents the maximum possible pixel value of the image and \gls{mse} is given by:
\begin{equation}
\text{MSE} = \frac{1}{N_1N_2} \sum_{i=1}^{N_1} \sum_{j=1}^{N_2} \left( I(i,j) - \hat{I}(i,j) \right)^2,
\end{equation}
with $N_1 \times N_2$ being the dimensions of the image.

The \gls{ssim} evaluates the similarity between two images based on luminance, contrast, and structural information. It is calculated as:
\begin{equation}
\text{SSIM}(x, y) = \frac{(2\mu_x \mu_y + C_1)(2\sigma_{xy} + C_2)}{(\mu_x^2 + \mu_y^2 + C_1)(\sigma_x^2 + \sigma_y^2 + C_2)},
\end{equation}
where $\mu_x$ and $\mu_y$ are the mean values, $\sigma_x^2$ and $\sigma_y^2$ are the variances, and $\sigma_{xy}$ is the covariance of the original and reconstructed images. $C_1$ and $C_2$ are constants to stabilize the division. \gls{ssim} values range from -1 to 1, with higher values indicating greater structural similarity between images.

\subsubsection{Transmission Performance Metrics}

Latency is defined as the time delay between the initiation of image transmission and its successful reception and reconstruction at the receiver. However, in our experiments, latency was quantified by the number of transmission slots required to transmit the image, excluding encoding and decoding times. We further break down latency metrics into:
\begin{itemize}
    \item Average Waiting Time ($T_{\text{avg}}$): The mean time delay (in ms) for image transmission.
    \item 99.9th Percentile Waiting Time ($T_{99.9\%}$): The time below which 99.9\% of image transmissions are completed, critical for delay-sensitive applications.
\end{itemize}

Throughput is measured as the effective data rate achieved during transmission, influenced by the modulation scheme and channel conditions. We use Megapixels per second (Mpps) as the unit, where each pixel is counted only once to reflect the actual image area transmitted. This metric ensures that multiple bits transmitted to refine the same pixels do not inflate the throughput measurement.

Reliability assesses the accuracy and integrity of the transmitted image data. In this study, reliability is evaluated based on the fidelity of image reconstruction as indicated by \gls{psnr} and \gls{ssim} values under varying channel conditions. Additionally, reliability encompasses the successful completion of image transmission without significant degradation, especially under poor \gls{snr} conditions.

\subsection{Implementation Details}
\subsubsection{\gls{vqgan}-based model}
For the \gls{vqgan}-based model, codebook generation and clustering are critical for progressive transmission. K-means clustering was employed using FAISS \cite{douze2024faiss} for efficient clustering on embeddings. Codebooks were generated for bits per index ($bpi$) ranging from 8 to 16 with a maximum of 100 iterations for the clustering algorithm and Euclidean distance (L2 norm) as the distance metric. Untimately, $M_{\text{stages}} = 10$ residual codebooks were trained, each with $bpi = 8$ (256 codewords), to support progressive residual transmission in the \gls{vqgan}-based model.
For the \gls{vqgan}-based model, progressive transmission is achieved by transmitting indices corresponding to coarse codebooks first, followed by finer codebooks. The receiver integrates these indices to incrementally refine the image reconstruction.

\subsubsection{Progressive Transmission Implementation}

In the hyperprior-based model, we implemented progressive transmission by masking feature maps based on their importance, as detailed in Section IV. The receiver progressively reconstructs the image as more feature maps are received.

\subsection{Experimental Procedure}

The experimental workflow was designed to systematically evaluate both the hyperprior-based and \gls{vqgan}-based models under identical channel conditions. The procedure is as follows:

\begin{enumerate}
    \item Data Preparation: The Kodak dataset images were preprocessed to match the input requirements of the models.
    \item Codebook Generation: For the \gls{vqgan}-based model, codebooks were generated using k-means clustering and residual quantization as described above.
    \item Channel Simulation: Simulated Rayleigh fading channels with varying \gls{snr} levels were generated for transmission simulations.
    \item Transmission Simulation: Each image was subjected to transmission over the simulated channels. The bit budget $N_{\text{bits}}$ for each transmission slot was determined based on the current \gls{snr} and selected modulation scheme.
    \item Progressive Decoding: The receiver progressively reconstructed images by integrating newly received data with previously received data, allowing for incremental improvements in image quality.
    \item Metric Collection: At each stage of progressive decoding, \gls{psnr}, \gls{ssim}, latency, and throughput metrics were recorded to evaluate performance.
    \item Result Aggregation: Results were averaged over all images in the dataset and multiple channel realizations to obtain statistically significant performance evaluations.
\end{enumerate}



\section{Results and Discussion}

In this section, we present experimental results evaluating the performance of our proposed adaptive and progressive image transmission pipeline over simulated Rayleigh fading channels. We analyze image quality, latency, throughput, and the effectiveness of our adaptive strategies. We compare the performance of the hyperprior-based and \gls{vqgan}-based models under various channel conditions and bit budgets, and discuss the implications of our findings for wireless communications.





\begin{figure}[t]
    \centering
    \includegraphics[width=\linewidth]{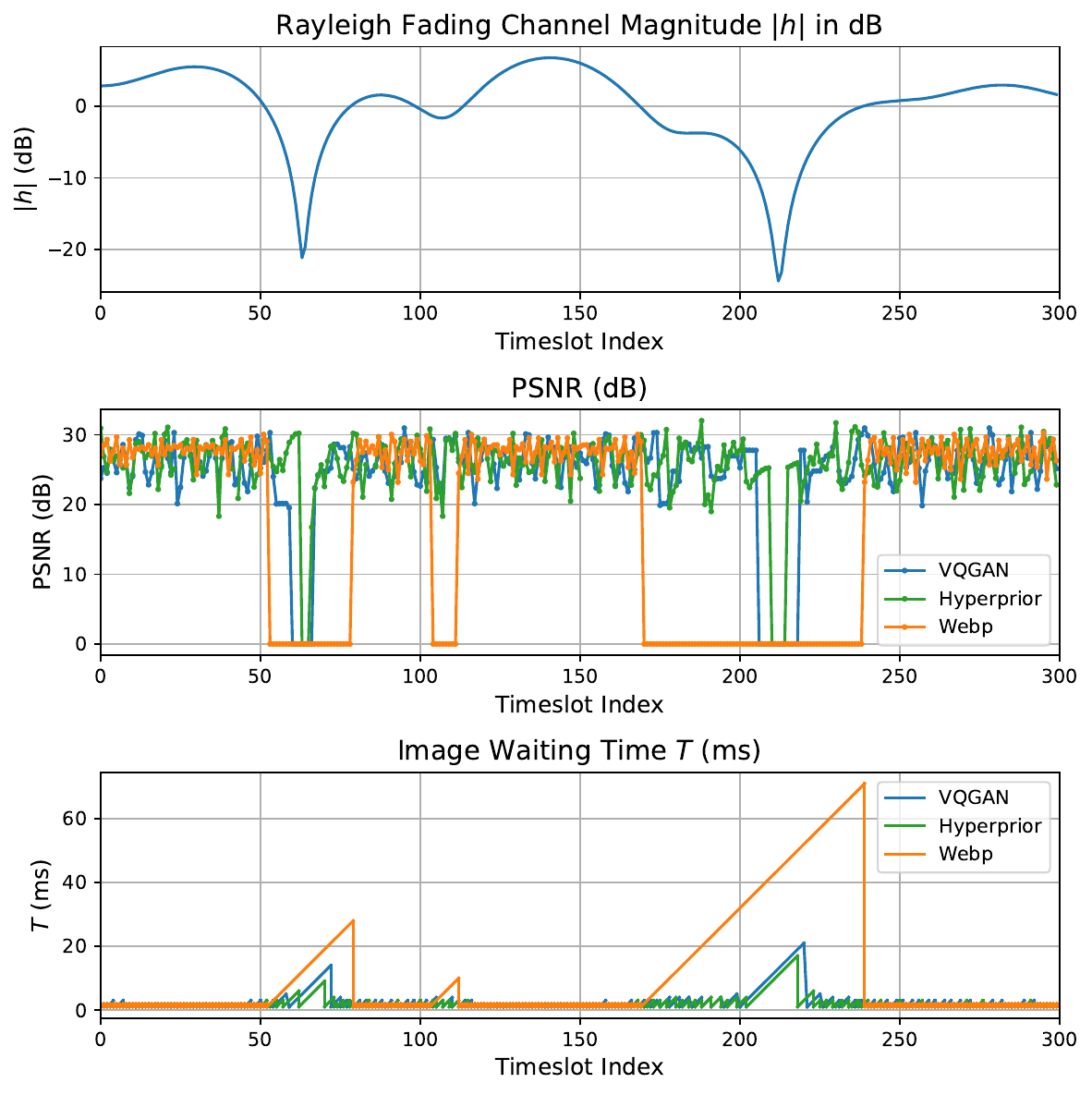}
    \caption{\small{A 300 ms snapshot of the fading channel magnitude $|h|$, PSNR of the transmitted images for the three models, and waiting time $T$ of the image transmission system for progressive-hyperprior, progressive VQGAN, and adaptive WebP models over the snapshot.}}
    \label{fig:fading_channel}
\end{figure}

\subsection{Experimental Results}

Table~\ref{tab:kodak_results_all_snr} illustrates the performance of our proposed models across various \gls{snr} values, providing a comprehensive view of throughput, latency, and image quality under realistic channel conditions. Each model is configured to maximize its performance within its respective architecture. The progressive-hyperprior model is designed to transmit up to 32 feature maps, balancing between resolution and minimal latency. progressive-VQGAN operates with 10 residual quantization stages (\(M_{\text{stages}}=10\)), aiming for higher fidelity under limited channel conditions. Finally, adaptive WebP dynamically adjusts quality factors (here from 1 to 4) to utilize available channel bandwidth effectively, optimizing image quality.

A snapshot of the Rayleigh fading channel magnitude $|h|$ over a timespan of 300 ms is presented in Fig.~\ref{fig:fading_channel}. This plot illustrates the temporal variations in channel conditions, which directly impact the image quality and transmission latency of each model. As observed, significant drops in channel magnitude correspond to stopped transmission for adaptive WebP, where it fails to deliver even the lowest quality image (quality factor = 1). In contrast, the progressive transmission schemes continue to function under these challenging channel conditions. This resilience is crucial for applications requiring consistent image quality under varying channel conditions.

\begin{table*}[htbp]
    \centering
    \caption{\small{Kodak Test Results for Proposed Models Across SNR Values.}}
    \label{tab:kodak_results_all_snr}
    \setlength{\tabcolsep}{4pt} 
    \small 
    \begin{tabular}{@{}ccS[table-format=3.2]S[table-format=2.2]cS[table-format=3.2]S[table-format=3.2]@{}}
        \toprule
        \multirow{2}{*}{SNR (dB)} & \multirow{2}{*}{Method} & \multicolumn{1}{c}{Throughput} & \multicolumn{1}{c}{PSNR} & \multicolumn{1}{c}{SSIM} & \multicolumn{1}{c}{$T_{avg}$} & \multicolumn{1}{c}{$T_{99.9\%}$} \\
        & & {Mpps} & {dB} &  & {ms} & {ms} \\
        \midrule
        \multirow{3}{*}{-10} 
            & Adaptive Webp          & 0.00    & \multicolumn{1}{c}{-}        & \multicolumn{1}{c}{-}     & \multicolumn{1}{c}{-}      & \multicolumn{1}{c}{-}      \\
            & Progressive-VQGAN      & 4.88    & \textbf{25.64} & \textbf{0.71} & 63.14  & 272.00 \\
            & Progressive-Hyperprior & \textbf{18.00} & 24.99   & 0.68  & \textbf{21.52} & \textbf{108.00} \\
        \midrule
        \multirow{3}{*}{-5} 
            & Adaptive Webp          & 0.00    & \multicolumn{1}{c}{-}        & \multicolumn{1}{c}{-}     & \multicolumn{1}{c}{-}      & \multicolumn{1}{c}{-}      \\
            & Progressive-VQGAN      & 34.13   & \textbf{25.83} & \textbf{0.72} & 11.14  & 94.00  \\
            & Progressive-Hyperprior & \textbf{66.38} & 25.28   & 0.69  & \textbf{6.62}  & \textbf{60.00} \\
        \midrule
        \multirow{3}{*}{0} 
            & Adaptive Webp          & 74.63   & \textbf{27.47} & \textbf{0.75} & 5.37   & 265.00 \\
            & Progressive-VQGAN      & 101.25  & 26.20   & 0.73  & 4.31   & 34.00  \\
            & Progressive-Hyperprior & \textbf{160.50} & 25.94   & 0.72  & \textbf{3.33}  & \textbf{21.00} \\
        \midrule
        \multirow{3}{*}{5} 
            & Adaptive Webp          & \textbf{454.50} & \textbf{27.63} & \textbf{0.76} & \textbf{1.82} & 71.00 \\
            & Progressive-VQGAN      & 205.13  & 26.31   & 0.73  & 2.67   & 16.00  \\
            & Progressive-Hyperprior & 307.88  & 26.35   & 0.73  & 2.22  & \textbf{12.00} \\
        \bottomrule
    \end{tabular}
\end{table*}

\subsection{Analysis of Results}

The experimental data offers insights into each model's strengths in terms of latency, image quality, and throughput across diverse \gls{snr} settings.

The Progressive-Hyperprior model consistently outperforms the other models in minimizing latency across all \gls{snr} conditions. At low \gls{snr} levels (-10, -5, and 0 dB), it achieves the lowest average waiting time \(T_{\text{avg}}\) and 99.9th percentile waiting time \(T_{99.9\%}\), demonstrating its effectiveness for real-time applications where speed is essential. Notably, even at 5 dB, where Adaptive WebP briefly offers a slightly lower \(T_{\text{avg}}\), Progressive-Hyperprior maintains a strong lead in 99.9th percentile latency, supporting its robustness under both challenging and favorable conditions for delay-sensitive applications.

This superior latency performance is primarily due to the progressive-hyperprior model's design, which bases its progressive decoding on compact feature maps. These compact feature maps enable the transmission of a granular and small number of bits, even in poor channel conditions.

Image quality, as measured by \gls{psnr} and \gls{ssim}, varies significantly with \gls{snr} and model choice:
Adaptive WebP yields the highest quality metrics (PSNR and SSIM) at higher \gls{snr} levels (0 dB and 5 dB), achieving peak fidelity in ideal channel conditions. However, it cannot transmit at -10, -5, and 0 dB due to channel capacity constraints. In contrast, progressive-VQGAN excels at lower \gls{snr} levels, outperforming progressive-hyperprior in terms of PSNR and SSIM. Its architecture is resilient to adverse conditions, providing higher image quality when Adaptive WebP fails to transmit. This suggests progressive-VQGAN’s suitability for scenarios demanding quality retention under limited bandwidth.

Throughput performance underscores the progressive-hyperprior’s efficiency:
At low to moderate \gls{snr} (-10 to 0 dB), progressive-hyperprior achieves the highest throughput, indicating optimal use of available channel capacity even under stringent conditions. This throughput advantage makes it suitable for applications where maximizing data transfer is essential despite poor channel quality. At 5 dB, Adaptive WebP surpasses other models in throughput, taking advantage of the higher quality factor options available at improved SNR. This transition illustrates Adaptive WebP’s capacity to exploit good channel conditions effectively but also underscores its dependency on sufficient bandwidth.

\begin{figure}[h!]
    \centering
    \begin{subfigure}{0.5\linewidth}
        \centering
        \includegraphics[width=\textwidth]{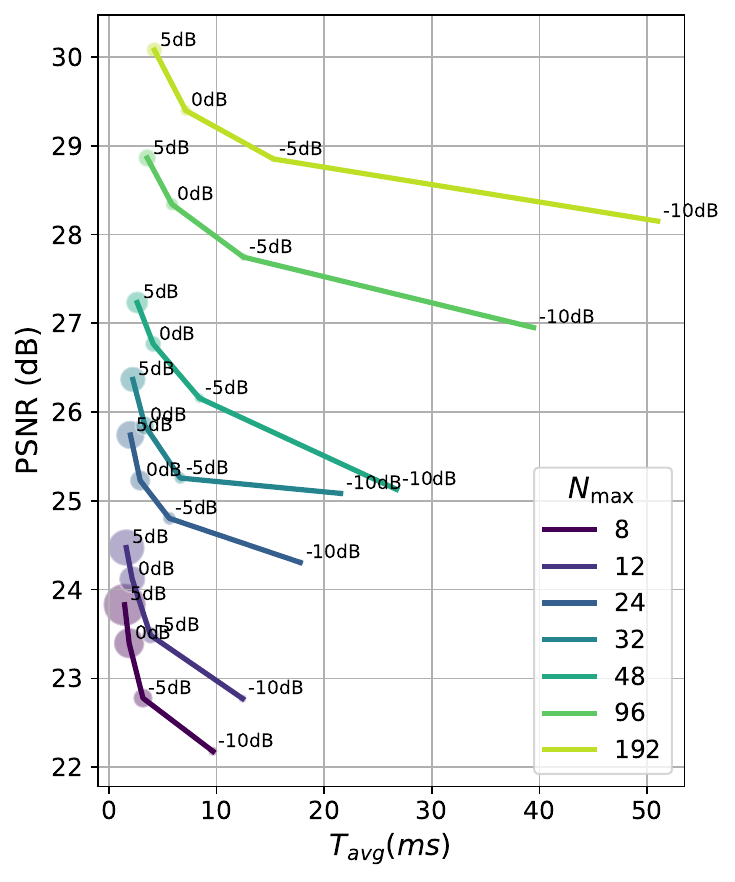}
        \caption{PSNR (dB) vs. $T_{avg}$ (ms)}
    \end{subfigure}\hfill
    \begin{subfigure}{0.5\linewidth}
        \centering
        \includegraphics[width=\textwidth]{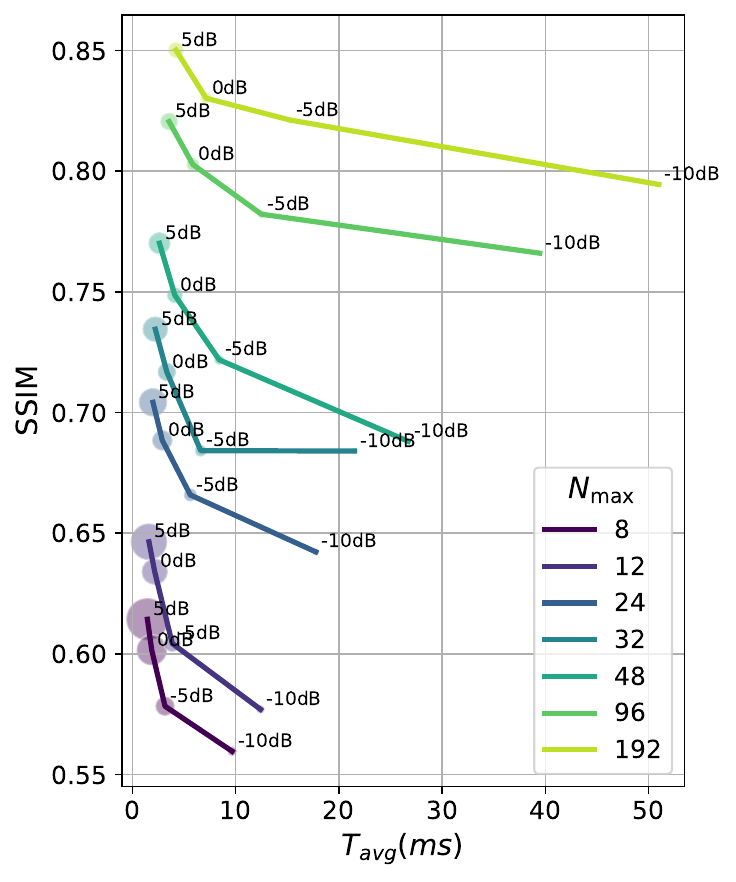}
        \caption{SSIM vs. $T_{avg}$ (ms)}
    \end{subfigure}
    \caption{Visualization of $T_{avg}$ (ms) and performance metrics for the hyperprior model: (a) PSNR (dB) and (b) SSIM. The size of the circles represents throughput, with annotations on the circles indicating the channel SNR. By adjusting $N_{\text{max}}$, which determines the number of feature maps to be transmitted for each image, we can observe the impact on PSNR and SSIM, reflecting the tradeoff between reliability, latency, and throughput in varying channel conditions.}
    \label{fig:ablation_N_max}
\end{figure}

Figure \ref{fig:ablation_N_max} illustrates the impact of the parameter $N_{\text{max}}$ on the tradeoff between throughput, reliability (PSNR and SSIM), and latency for the hyperprior model. The parameter $N_{\text{max}}$ specifies the total number of feature maps to be transmitted for each image, enabling us to adapt the model's performance to different channel conditions. By examining this figure, we can select the appropriate value of $N_{\text{max}}$ to achieve a desired tradeoff. For instance, if the target performance requirements are SSIM $> 0.75$ and PSNR $> 27$ dB, then $N_{\text{max}} = 96$ and $N_{\text{max}} = 192$ both meet these criteria. However, since $N_{\text{max}} = 96$ allows for higher throughput (transmitting more images per second), it would be the preferred choice. This analysis helps the selection of $N_{\text{max}}$ based on specific performance and throughput priorities. 

\subsection{Discussion}

The findings emphasize the adaptability of each model to specific use-case demands, showcasing the trade-offs between latency, quality, and throughput. Progressive-hyperprior is ideal for delay-sensitive applications, minimizing latency while maintaining throughput across various SNR levels. Its selective transmission of feature maps allows it to remain functional under poor conditions, a trait that could benefit applications such as interactive video streaming or remote sensing. Progressive-VQGAN’s robustness in low \gls{snr} conditions makes it advantageous where quality is a priority and latency is secondary. Its vector quantization approach allows reliable image reconstruction, reducing the need for channel coding (hence encoding and decoding time) lowering computational load. Progressive-VQGAN’s performance under degraded channels highlights its utility for applications that prioritize visual fidelity under bandwidth constraints. Performing best in high-quality conditions (5 dB), Adaptive WebP achieves the highest image quality and throughput in such environments. However, its dependency on channel capacity restricts its usability in lower \gls{snr} scenarios. Adaptive WebP’s efficiency in optimal conditions positions it as an ideal choice for high-definition media streaming where channel conditions are controlled or consistently high.

\subsection{Implications for Wireless Communications}

The adaptability and progressive nature of these models provide valuable tools for optimizing wireless communication systems, particularly in variable or constrained environments. By tailoring the transmission approach to the channel’s real-time conditions, our framework enables targeted optimization for latency, quality, or throughput as the application requires.

Real-time and low-latency applications can benefit from progressive-hyperprior’s latency performance, which supports immediate data access needed for interactive services. For scenarios where image quality is essential, particularly under stable and high \gls{snr} conditions, Adaptive WebP or progressive-VQGAN can achieve high PSNR and SSIM values, making them ideal for applications where visual clarity is crucial. The robustness of progressive-VQGAN without channel coding reduces processing demands, beneficial in resource-limited environments or applications where computational efficiency is critical.


\section{Conclusion}

In this paper, we proposed an adaptive and progressive image transmission pipeline for wireless communication systems using \gls{lic} architectures. By leveraging hyperprior-based and \gls{vqgan}-based models, our approach effectively balances reliability, throughput, and latency in dynamic wireless channels.

Our experiments over simulated Rayleigh fading channels demonstrated that the proposed pipeline enhances image transmission performance under varying channel conditions. The progressive-hyperprior model achieved the lowest latency, making it suitable for delay-sensitive applications, while the Progressive-\gls{vqgan} model provided superior image quality in challenging channel conditions without the need for channel coding. In contrast, traditional methods like Adaptive WebP showed limitations under poor channel conditions.

These findings highlight the advantages of integrating advanced \gls{lic} architectures into wireless communication protocols. Our adaptive and progressive transmission strategy enables real-time image transmission that is resilient to channel variations. The integration of learning-based compression techniques represents a significant step toward meeting the demands of next-generation communication networks.



Our adaptive and progressive image transmission pipeline effectively balances the trade-offs between reliability, throughput, and latency in wireless communication systems. By utilizing advanced \gls{lic} architectures and tailoring the transmission strategy to real-time channel conditions, we have demonstrated substantial improvements over traditional methods. The ability to maintain robust image transmission even under challenging conditions opens up new possibilities for a wide range of applications that rely on timely and accurate visual data. The integration of learning-based compression techniques into wireless communication protocols represents a significant step toward meeting the demands of next-generation communication networks.

\section*{Acknowledgments}
This research is funded by the Smart Networks and Services Joint Undertaking (SNS JU) under the European Union’s Horizon Europe research and innovation program (Grant Agreement No. 101139194, 6G-XCEL), as well as the Horizon Europe program under the MSCA Staff Exchanges 2021 (Grant Agreement No. 101086218, EVOLVE project).

\bibliographystyle{IEEEtran}
\bibliography{main}
\end{document}